\documentclass[12pt,preprint]{aastex}  
\def\magsqarcsec{mag$/$\raisebox{-0.4ex}{\hbox{$\Box^{\prime\prime}$}\,}} 
\newcommand{\oversim}[2]{\protect{\mbox{\lower0.5ex\vbox{%
   \baselineskip=0pt\lineskip=0.2ex   
   \ialign{$\mathsurround=0pt #1\hfil##\hfil$\crcr#2\crcr\sim\crcr}}}}}    
\newcommand{\simless} {\mbox{$\,\mathrel{\mathpalette\oversim<}\,$}}
\begin{document}   
\shorttitle{The Tidal Stream of NGC\,5907}   
\shortauthors{Mart\'{i}nez-Delgado et al.}   
\title{The ghost of a dwarf galaxy: \\
  fossils of the hierarchical formation of the nearby spiral galaxy
  NGC\,5907}
\author{David Mart\'{i}nez-Delgado\altaffilmark{1,2,7}, Jorge
  Pe\~narrubia\altaffilmark{3}, R.Jay
  Gabany \altaffilmark{4}, Ignacio Trujillo\altaffilmark{1,7},\\ Steven R. Majewski \altaffilmark{5}, M. Pohlen
  \altaffilmark{6}}
\altaffiltext{1} {Instituto de Astrof\'isica de Canarias, La Laguna, Spain}
\altaffiltext{2} {Max-Planck Institut f\"ur Astronomie, Heidelberg, Germany}
\altaffiltext{3} {University of Victoria, Canada} 
 \altaffiltext{4}
             {Blackbird Observatory, New Mexico, USA}
\altaffiltext{5} {Department
  of Astronomy, University of Virginia, USA} 
\altaffiltext{6} {Cardiff
  University, School of Physics \& Astronomy, Cardiff, UK}
 \altaffiltext{7}
             {Ram\'on y Cajal Fellow}

\begin{abstract}   

We present  an extragalactic perspective of an extended
stellar tidal stream wrapping around the edge-on, spiral galaxy NGC\,5907.
Our deep images reveal for the first time a large scale complex of
arcing loops that is an excellent example of how a low-mass satellite
accretion can produce an interwoven, rosette-like structure of debris
dispersed in the halo of its host galaxy. The existence of this
structure, which has probably formed and survived for several Gigayears, 
confirms
that halos of spiral galaxies in the Local Universe may still contain a
significant number of galactic fossils from their hierarchical
formation.

To examine the validity of the external accretion scenario, we present
N-body simulations of the tidal disruption of a dwarf galaxy-like
system in a disk galaxy plus dark halo potential that demonstrate that
most of the observed tidal features observed in NGC\,5907 can be
explained by a single accretion event. Unfortunately, with no
kinematic data and only the projected geometry of the stream as
constraint, the parameters of our model are considerably degenerate
and, for now, must be considered illustrative only.

Interestingly, NGC\,5907 has long been considered a prototypical example of a
warped spiral in relative isolation. The presence of an extended tidal stream
challenges this picture and suggests that the gravitational perturbations
induced by the stream progenitor must be considered as a possible cause for the warp. 
The detection of an old, complex tidal stream in a nearby galaxy with
rather modest instrumentation points to the viability of surveys to
find extragalactic tidal substructures around spiral galaxies in the
Local Volume ($<\!$ 15 Mpc) --- with the prospect of obtaining a
census with enough statistical significance to be compared with
cosmological simulations.
\end{abstract}   
\section{Introduction}\label{int}   
The Cold Dark Matter ($\Lambda$CDM) paradigm predicts that stellar halos surrounding
large spiral galaxies formed through the accretion and tidal disruption of
satellite galaxies, a notion previously postulated \citep{Searle1978} on
empirical grounds from the character of stellar populations found in our own
Milky Way halo. Simulations have shown that the fossil records of those merger
events may be detected nowadays in the form of coherent stellar structures in
the outer regions of the host galaxies. The most spectacular of these fossil
structures are probably the formation of dynamically cold stellar streams,
which are similar to long rivers of stars (and likely dark matter; \cite{Penarrubia2008b}) tidally
stripped from a disrupting dwarf galaxy and wrapped around the host galaxy disk,
roughly tracing the orbit of the progenitor satellite. The now well-studied
Sagittarius (Sgr) tidal stream surrounding the Milky Way \citep{Ibata2001b,
  Majewski2003, Martinez-Delgado2004} and the giant stream in Andromeda galaxy
\citep{Ibata2001a} are archetypes in the Local Group of these kinds of
fossilized satellite galaxy mergers and provide obvious support for the
scenario that tidally disrupted dwarf galaxies are an important contributor to
the stellar halos of disk galaxies.

Unfortunately, while the Milky Way and even M31 streams are close enough to be
resolved into individual stars that can be studied in detail, these nearby
streams can also span vast areas of sky, which currently presents particular
challenges to their study. In contrast, tidal streams around more distant
systems can be imaged in commensurately smaller fields of view, and --- at
least for those streams seen in diffuse light --- with equal contrast, given
the constancy of surface brightness with distance. However, over the past
decade, only a few cases of confirmed stellar tidal streams have been detected
in disk galaxies outside the Local Group \citep[e.g., ][and references
  therein]{Malin1997, Forbes2003, Pohlen2004}, in contrast with the increasing
number of tidal structures detected in the Milky Way and M31 in recent years.
The presence of a multitude of tidal streams, arcs, shells and other irregular
structures possibly related to ancient merger events 
 in the deep panoramic view of the Andromeda halo  \citep{Ibata2007}
 reveals the level of stellar sub-structure that might be 
 present in the halos of nearby external spiral galaxies.

Extragalactic streams also offer a unique perspective to address some
interesting questions: Is the abundant number of stellar streams exceptional
in the Local Group or are tidal streams as common as has been predicted by
current cosmological models?  What can we learn about the dark-matter halos of
the host galaxies from these streams?  Can we use these systems to study the
star formation history of the disrupting dwarf galaxies?  The frequency of
streams, their stellar population and orbital properties provide important
clues to the nature of the stream progenitors, lend insights into the
underlying gravitational potential and structure of the massive dark halo
they inhabit, and constrain hierarchical models of galaxy formation favored by
current cosmological models. In that respect, an interesting parameter that may possibly be constrained
through the detection of tidal streams in external galaxies is the flattening
of the dark matter halos surrounding the host galaxies. The spatial and
kinematic distributions of stream particles are fairly sensitive to this
quantity \citep[e.g.][and references therein]{Johnston2002, Helmi04,
  Johnston2005, Law2005, Penarrubia2006} and the shapes and orientations of
dark matter halos with respect to the disk axi-symmetry plane reflect the nature
of dark matter \citep[see, e.g.][for shape estimates for cold, self-interacting and hot dark
matter models, respectively]{Dubinsky2001, Yoshida2000, Dave2001}. For example, $\Lambda$CDM models predict that dark matter
halos are triaxial, with minor axes perpendicular to the disk plane. On average the minor-to-major axis ratio near
the center is $c/a\simeq 0.7$ \citep{Hayashi2007}, although halos may become
considerably rounder after the baryonic galactic components have formed
\citep{Kazantzidis2004} or if dark matter is warm \citep{Biermann2007}. Thus,
using tidal streams in external galaxies might be an ideal tool to obtain a
significant statistical sample of flattening values that can be compared
against predictions from cosmological models.

 These arguments suggest that a systematic survey
for tidal streams around extragalactic systems would provide a new way to
explore galaxy structure and evolution, with direct constraints on
cosmological models. Undertaking such a tidal stream survey 
was urged by \cite{Johnston2001} in their comprehensive discussion of the
appearance and detectability of extragalactic tidal streams. Their analysis
suggests that a survey of 100 parent galaxies reaching to a surface brightness
of 30 mag arcsec$^{-2}$ would reveal many tens of tidal features; a slightly
higher estimate of about one per galaxy is obtained in their most recent
reanalysis of the problem \citep{Bullock2005}. These authors also point out
the advantage of exploring thin, edge-on disk systems because of the favorable
orientation for detecting low surface brightness features against light of the
parent system and the ability to assess limits on the mass ratio of the
satellite and parent galaxy by whether the disk of the parent appears to be
disturbed.

\cite{Johnston2001} estimated that a search strategy designed to obtain
reliable photometry along the stream length would require about ten times
longer exposure times that those aimed at detection only. We adopted this
 approach of striving for deep images to detect the full path of the
tidal stream, which, in the absence of kinematic data, is the main input for
comparison with N-body simulations (see Sec.\ref{nbody}). Using this strategy,
 we started a pilot survey on some nearby galaxies which
has proven the concept with the discovery of a giant stellar tidal stream
around the spiral galaxy NGC\,4013 \citep{Martinez-Delgado2008} and the
detection of the faintest parts of the extended disk of the nearby spiral
galaxy M\,94 (Trujillo et al. 2008, in preparation).

With the hope of improving our understanding of the known tidal stream in
NGC\,5907, discovered by \cite{Shang1998}, and to demonstrate the sensitivity
of the small aperture telescope used in our survey (see Sec.\ref{data}), we
reobserved this system as a commissioning target.
In this paper we present our very deep image of NGC\,5907 that reveals
additional loops of debris in the halo (see Sec.\ref{sec:shape}) beyond those
reported by \cite{Shang1998}.
N-body simulations, described in Sec.\ref{nbody} show that most of the newly
found features can be explained as pieces of a spectacular, multiply-wrapped
stream of debris from the tidal disruption of a single companion galaxy. 
We also discuss in Sec.\ref{discussion} the applications of such surveys to glean a deeper
theoretical understanding of galaxy formation and evolution, beyond placing
strong constraints on present cosmological scenarios. This foray into a more
systematic look for tidal streams in the nearby universe has yielded promising
results that encourage more aggressive attention to this new way of
understanding galaxy structure and evolution as advocated by the above
theoretical studies.
\section{Target, Observations and Data Reduction}  
\subsection{The Target: NGC\,5907}
NGC\,5907 is a nearby \citep[14 Mpc;][]{Zepf2000}, edge-on, Sc galaxy, very
similar (e.g., type, absolute magnitude, and rotation speed) to the Milky Way.
NGC\,5907 has frequently been the target of deep surveys (radio, optical and
near-infrared) because it is an apparently undisturbed, almost edge-on disk
useful for exploring the properties of spiral disks, determining the
properties of dust and gas in disk galaxies, testing for the prevalence of
luminous thick disks and halos, searching for warps, and exploring the
strength of dark matter in disks \citep[e.g.][]{Sancisi1976, vdK1981,
  Casertano1983, Skrutskie1985, Sasaki1987, Barnaby1992, Morrison1994,
  Sackett1994, Lequeux1996, Lequeux1998, Rand1996, Dumke1997, Rudy1997,
  James1998, Xilouris1999, Zheng1999}.
Although VLA HI maps of this galaxy showed that it has a large-scale
warped gas disk \citep{Sancisi1976}, NGC\,5907 was long considered to
be a prototypical example of a warped disk galaxy existing in relative
isolation since the identification of a neighbouring,
interacting galaxy was lacking.
However, deep imaging of NGC\,5907 by \cite{Shang1998} has revealed a
very faint, elliptically-shaped ring around the disk, and this has
been interpreted to be the remnant of a tidally disrupted dwarf
satellite galaxy, an analog to the debris stream created by the Milky
Way-Sgr interaction, although at higher surface brightness.
\subsection{Observation and Data Reduction} 
\label{data}

We have obtained deep optical images of NGC\,5907  with the 0.5-meter
Ritchey-Chretien telescope of the BlackBird Remote Observatory (BBRO)
 situated in the Sacramento Mountains (New Mexico, USA). We used a Santa
 Barbara Instrument Group (SBIG) STL-11100 CCD camera, 
which yields a large field of view (27.7$\arcmin$ $\times$ 18.2$\arcmin$) at a plate scale of 0.45 
arcseconds pixel$^{-1}$. The image set consists of multiple deep exposures 
with a non-infrared, clear luminance ($3500<\lambda<8500$), a red, a green and a blue filter from the SBIG Custom Scientific filter set.
Table \ref{obslog} provides a  summary of the images collected during
different dark sky observing runs during the period June through 
August 2006. Column 3 refers to the total
 exposure time of the co-added images from each filter obtained with each run.

The science images were reduced using the standard procedures for bias
correction and flat-fielding. A master dark and bias frame was created by
combining 10 dark sub-exposures each produced at the same exposure length and
camera temperature settings used for the luminance and the filtered images. A
master flat was produced by combining 10 separate sky flat exposures for each
filter. The red, green and blue filtered exposures were separately combined
(using a median procedure) to produce red, green and blue master images that
represented the total exposure time for each color filter. These color images
 were obtained to investigate the color of the structures surrounding the
 galaxy and to extend the overall exposure of the final luminance
 image.
 The master images
for each filter were subsequently summed to produce a synthetic luminance
image that represented the total exposure time of all the color filter
data. The synthetic luminance image was then combined  (using a median
procedure) with all the clear filtered luminance sub-exposures (see Table 1)
to increase the contrast of the diffuse light. The resulting final image thus
represents the sum of all available CCD exposures collected for this project,
with an accumulated exposure time of 11.35 hours (including 5.75 hours in the clear luminance filter).

 To enhance the faint structures around  NGC\,5907 and disentangle their true
 appearance from image noise effects, the contrast and detail of the summed
 image was optimized  using two well-known image processing techniques:
i)a histogram equalization of the image  using a non-linear transfer
function\footnote{see
  http://http://homepages.inf.ed.ac.uk/rbf/HIPR2/histeq.htm and references
  within};
and ii) Gaussian blur filtering \citep[]{Davies1990,
  Haralick1992}.
 The first method is very effective to
reveal very faint features in images that contains a large dynamical range,
 as it is the case of a spiral galaxy with  a very bright, compact stellar
 disk  surrounded by a
 very faint tidal stream. With the purpose of suppressing the bright portions
 of the image and intensifying the fainter parts of the stream, histogram
 equalization employs a monotonic, non-linear mapping which re-assigns the
 intensity values of pixels in the input image such that the output image
 contains an uniform distribution of intensities. In our case, we used  an
 iterative process  that involves several passes of a S-shaped histogram
operator to the total image. 
While this method was successful in revealing the presence of the surrounding
faint structure, it also resulted in 
the over saturation of contrast and significantly reduced intensity variances
within the stellar stream. 
Once the image was contrasted, a low-pass Gaussian filter  with a radius of 2 pixels was applied
 to reduce the low level noise contaminating the image.
 The resulting noise-filtered image is shown in Fig.\ref{BBROimage}.

As discussed above, our search technique was designed to very clearly reveal
the position and morphology of faint structures (for example, by combining all the
available images obtained in different filters with the clear luminance deep
images), but does not permit accurate photometry. Therefore, follow-up
photometric observations are needed to
 measure accurate surface brightnesses and to estimate the total luminosity of
 the detected tidal debris (see Sec.~\ref{SB}).
 Furthermore, the use of an uncalibrated luminance filter prevents us from
 estimating the surface brightness
 limit that we reached in our deep probe of the NGC\,5907 halo in any
 particular, standard photometric band.
 However, the detection of obvious diffuse light structures fainter than those
 reported in previous studies 
(with a surface brightness as faint as $\Sigma_R \sim$ 27.7 mag arcsec$^{-2}$;
see Sect.~\ref{SB}), suggests
 that our images reach a surface brightness magnitude fainter than the $R=27$
 mag arcsec$^{-2}$ limit of the deep NGC\,5907 
image by  \cite{Morrison1994} and \cite{Sackett1994}, and the $R=28.7$ mag arcsec$^{-2}$ limited image by \cite{Zheng1999}.

Regardless, this result is a striking example of the scientific potential of
modest aperture telescopes (0.5-meter)
 operating under very dark skies for probing the diffuse structure of galaxies.

\section{The Tidal Stream(s) of NGC\,5907}  
\subsection{Observed Morphology of the NGC\,5907 System} 
\label{sec:shape}   
\subsubsection{Truncation and Optical Warp}

\begin{figure}[!]   
\vspace{15.2cm}   
\includegraphics{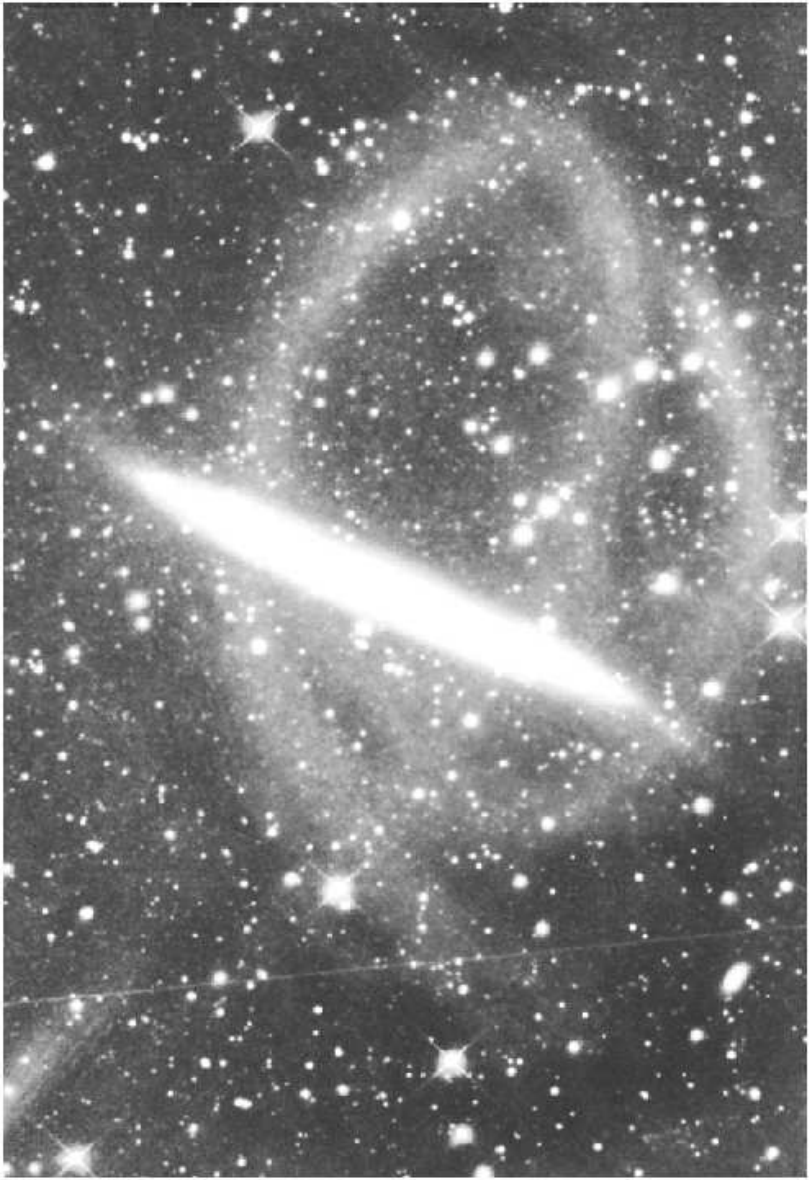}   
\caption{Image of NGC\,5907 obtained with the BBRO 0.5-meter telescope. The total
  exposure time of this image is 11.35 hours, co-adding all images obtained in this project (see Table
  \ref{obslog}). The image has dimensions of 18.2 $\times$ 27.7 arcmins, which, at
  the distance of NGC\,5907 is $\sim$ 75 $\times$ 115 kpc. For a better
  comparison with the N-body simulations given in Fig.~\ref{nbody}, this image
  is shown east up and north to the right. The linear diagonal feature in the lower left corner of the image is
  spurious light from deflection of a bright star off of the edge of the CCD
  chip. The faint halos surrounding the field stars are due to the pass of the Gaussian filter mentioned in Sec. \ref{data}.}
 
\label{BBROimage}   
\end{figure}

In this Section we focus on the outer, diffuse light properties revealed in
Figures \ref{BBROimage} and \ref{BBROimagePlabel}, but several comments about
the disk are worth pointing out briefly.

Our image shows that the disk of NGC\,5907 extends in diameter well
beyond where most previous images have tracked its disk light. The existence
of this light challenges previous claims that the disk is {\it sharply
  truncated} (in the sense of the galaxy's outer edge) at a nominal {\it
  cut-off} radius of about 6\arcmin ~from the center
\citep{vdK1981, Sasaki1987, Morrison1994}. From the original radial profiles 
\citep[][Fig.9 and Fig.7,
  respectively]{vdK1981,Sasaki1987} there is clearly a radial break feature
visible, however, it should be better described as a sharp change of slope than
a complete {\it cut-off}. This profile shape, well fitted with a broken
exponential function, is a relatively common feature of late-type spiral
galaxies \citep{Pohlen2006,Erwin2008}.

Figures \ref{BBROimage} and \ref{BBROimagePlabel} clearly show that NGC\,5907
has a stellar warp, which clearly extends after the nominal cut-off radius.
The observable faint radial extensions of the disk are strongly bent in the
same sense as the obvious HI warp found by \cite{Sancisi1976}, and are most
strongly visible on the northwest disk edge. Note, these same extended disk
features can also be seen in Fig.\ 2 of \cite{Shang1998}, and just about in
the near infrared image of \cite{Barnaby1992}.
Evidence for the warp in optical light was also previously seen in the optical
disk at smaller radii by \cite{Sasaki1987}, \cite{Florido1992}, and
\cite{Morrison1994} and was, of course, pointed out as well by
\cite{Shang1998}.  
\subsubsection{Diffuse Light Structures}

\begin{figure}[!]   
\vspace{17.2cm}   
\includegraphics{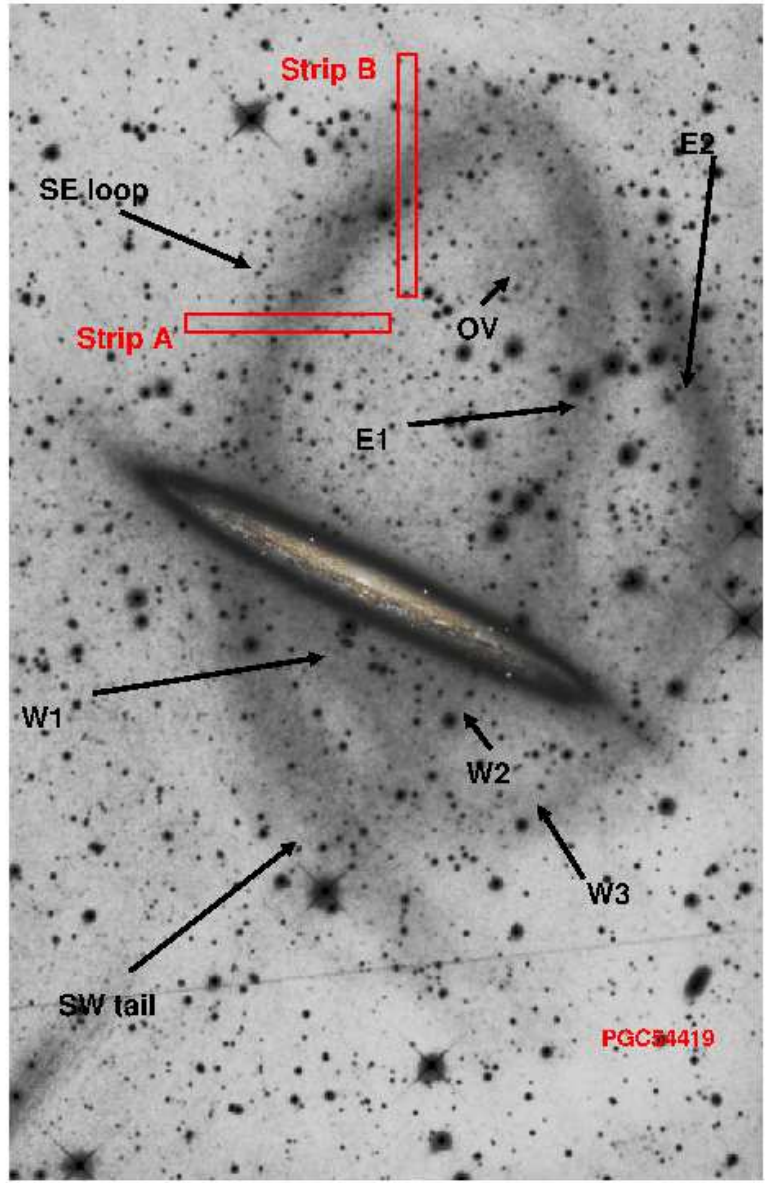}   
\caption{Identified photometric features detected in the image of NGC\,5907
  obtained with the BBRO 20-inch telescope and discussed in Sec.3.1. For reference, a colour  image obtained with the same telescope (see Table
  \ref{obslog}) has been superimposed on the saturated disk region of the  galaxy. The
  position of the two different strips used to measure the width of the stream
  (see Sec. 3.2) are also indicated by red rectangles. The yellow lines mark
  the position of the nominal {\it cut-off} radius (see Sec 2.2). For comparison
  purposes, north is to the right and east is up.}
\label{BBROimagePlabel}   
\end{figure}

The most striking features, and the primary focus of the present discussion,
is the newly revealed network of arcing loops of diffuse light around
NGC\,5907 that are likely the product of the classical satellite tidal
disruption process.  Our deep image (see Fig.\ref{BBROimagePlabel}) reveals that the
ring-shaped feature to the southeast, we refer to this here as the {\it SE loop},
discovered by \cite{Shang1998} is only the brighter part of a large-scale
complex of stellar debris arcs around NGC\,5907. We count about a
half-dozen other additional arc-like features in our image.  On the eastern
side, it is possible to detect the full extension of the SE loop, turning
around and falling toward the disk. This yields a vertical structure (labeled
E1 in Fig.\ref{BBROimagePlabel}) that crosses the galactic plane and possibly
extends to the other side of the galaxy.  In addition, an independent
structure (labeled E2), somewhat parallel to the SE-loop, can be seen to be
shifted about $\sim$ 10 $\arcmin$ to the northwest of the \cite{Shang1998}
structure and sweeping almost 180$^{\circ}$ around the center of NGC\,5907.

Although the accretion origin of this structure appears a reasonable
hypothesis, we must point out its extraordinary complexity --- especially on the
western side of NGC\,5907 where we find a faint, long tail extending $\sim$ 25 kpc from the
disk, labeled {\it SW tail} in Fig. \ref{BBROimagePlabel}, and previously
reported only 
in some studies \citep[see Fig.\ 1b in][]{Reshetnikov2000}.  From
south to north, three additional arcs are detected on the west side of the
disk (labeled W1, W2 and W3, respectively).  These features appear to
cross each other fairly closely to the disk major axis of NGC\,5907, which
makes it difficult to discern which of these arcs are the continuation of the
features observed on the eastern side, that is, the SE loop, E1, and E2.

For example, it is obvious that the SE loop extends to the western side, but
it becomes unclear which path it takes (e.g., SW tail or W1). On the other
hand, it is clear that the W3 arc is the westernmost continuation of the E2
loop. Interestingly, this stream piece turns around, crosses the W2 arc and
finishes in the main disk again, forming the W1 arc.  It is however more difficult to assess
whether the W2 arc (the most diffuse discernible structure,
highly contaminated by the diffuse light of the galaxy disk) or W1 could be the
western extension of the E1 tail.

\subsubsection{Evidence for a Single Coherent Structure}

With the image in Fig. \ref{BBROimagePlabel} alone it is impossible to conclude
whether the smaller features on the western side are the continuation of the
loops on the eastern side and so whether the whole set of these identified
tidal structures originate from a single or different accretion events.
In general, the largest features (e.g., the SE and E2 loops) have similar
shapes, angular lengths, radius of curvature, and separations from the center
of NGC\,5907.  In addition, all of the features (large and small) share a
similar projected width, $\sim$ 7 kpc \citep[consistent with the value
  reported by][see Sec.\ref{SB}]{Shang1998}, as well as a similar, rather
bright surface brightness (see Sec.\ref{SB}). These similarities all provide
circumstantial evidence that the various features may all come from a single,
old\footnote{Following \cite{Johnston2001}, the age of the trail here refers to the epoch when the
debris were unbound from the satellite core (see
Fig.~\ref{nbodyfig} ) and not to the age of its stellar population, that is
unknown.} stellar tidal stream --- similar to the multiply-wrapped Sgr dwarf
galaxy stream around the Milky Way. These structures provide a rather unique
and spectacular external view of the classical rosette-like, multi-loop
patterns expected for tidally disrupting dwarf galaxies seen in N-body
simulations \citep[such as those made of the Sgr system ---
  e.g., ][]{Johnston1995, Ibata1998, Martinez-Delgado2004, Law2005}.  There are
further analogies of the NGC\,5907 structures with the Sgr stream in that they
have a similar linear cross-sectional width (suggesting similarly massed
progenitors), apogalacticon and apparent orbital ellipticity (suggesting a
similar orbit), and surface brightness and length (suggesting a similar mass
loss rate and age of interaction). In Section \ref{nbody} we explore this
working hypothesis in detail and provide a comprehensive, illustrative model
that further supports the notion that a single, Sgr-like system could create a
substantial fraction of the newly found diffuse structures.

Interestingly, while the surface brightness of these loops does show some
modulation, nowhere among these features do we see a bright spot that could be
identified with a remaining dwarf satellite {\it core}.  Thus, our image provides
no definitive insights on the current position and final fate (that is,
whether or not it is completely disrupted by now) of the progenitor satellite
galaxy (since it could be hidden behind the disk).  However, the full
panoramic view of the NGC\,5907 system in Figure \ref{BBROimage} provides no
evidence for any association of the stellar streams with the dwarf irregular
galaxy  PGC\,54419 (a confirmed satellite of NGC 5907 situated at 36.9 kpc
projected distance from its center), as proposed by \cite{Shang1998}.
 We shall return to this subject
in Sec.\ref{discussion}.

\subsection{Diffuse Light: Halo or Stream?} 
\label{halo}   
As mentioned above, there have been a number of deep photometric studies of
NGC\,5907, and a popular point of discussion is the claimed detection and
properties of `` excess'', non-disk light away from and perpendicular to the
disk and commonly attributed to a diffuse luminous halo \citep{Sackett1994,
  Lequeux1996, Lequeux1998, Rudy1997, James1998, Zepf2000, Irwin2006}.  Figure
\ref{BBROimage} clearly shows how these previous studies may have been
affected by the coherent tidal stream structures, especially W1 and W2, which
lie along the southwestern minor axis at distances where such studies have
made the ``halo" detections.  Indeed, the very shape of the ``inner triangle"
of light formed by W1 and W2 can be discerned in the image analyzed by
\citet[see, e.g., their Fig.\ 1]{Lequeux1996}, and a similar asymmetry seen in
the photometric data by \cite{Sackett1994} may also be explained by these same
structures \citep[e.g., see the lower right panel of
  Fig.~\ref{BBROimagePlabel} in][]{Sackett1994}.

One wonders if the presence of these structures may also explain the
differences in the colors of the ``halo" light found on either side of the
disk shown in Figure 1 of \cite{Lequeux1998}.  \cite{Zheng1999} have
previously suspected that the presence of tidal debris light (particularly on
the western side) might be significantly contaminating the ``halo" light that
was previously reported; these authors concluded that, when combined with
residual scattered light from foreground stars, this debris effectively
reduces the need for a faint, extended halo. 
 We concur that indeed these previous studies were
probably strongly affected by the presence of the ``W'' features --- but of
course this {\it is} actually ``halo" light, it is just in the form of tidal
debris in what is obviously the active hierarchical formation of the NGC\,5907
halo. However, a low surface brightness halo
($\Sigma_R \sim$ 30-31 mag arcsec$^{-2}$) similar to M31 one could be present
(Fliri et al. 2008, in preparation).
\subsection{Estimating the Surface Brightness, Total Luminosity and Stellar Mass  of the NGC\,5907 Tidal Stream}  
\label{SB} 
Ideally, one would like to use the BBRO image to measure the surface
brightness of the observed stream. However, there are some concerns that
prevent us from extracting this information directly from this image.  First
of all, the image is the result of the combination of multiple images in
different filters with those taken using an extremely wide (luminance) filter to
maximize the photon collection. This makes a calibration of the stars present
in the BBRO image using the same stars observed through standard filters
non-trivial. Significant color corrections are expected.
In addition, the resulting background on the BBRO image is slightly patchy (see Fig. 1), and,
more importantly, the field of view is not large enough to allow us to properly
estimate the sky level.  For the above reasons we have decided to
adopt the value reported by \cite{Zheng1999} for the mean surface brightness
of the stream in the NE loop in the $R$-band --- i.e. 27.7 mag arcsec$^{-2}$ ---
and rely on the BBRO image only to determine the morphology, position and width
of the stream in the sky. 

\begin{figure}[!]   
\vspace{15.2cm}   
\includegraphics{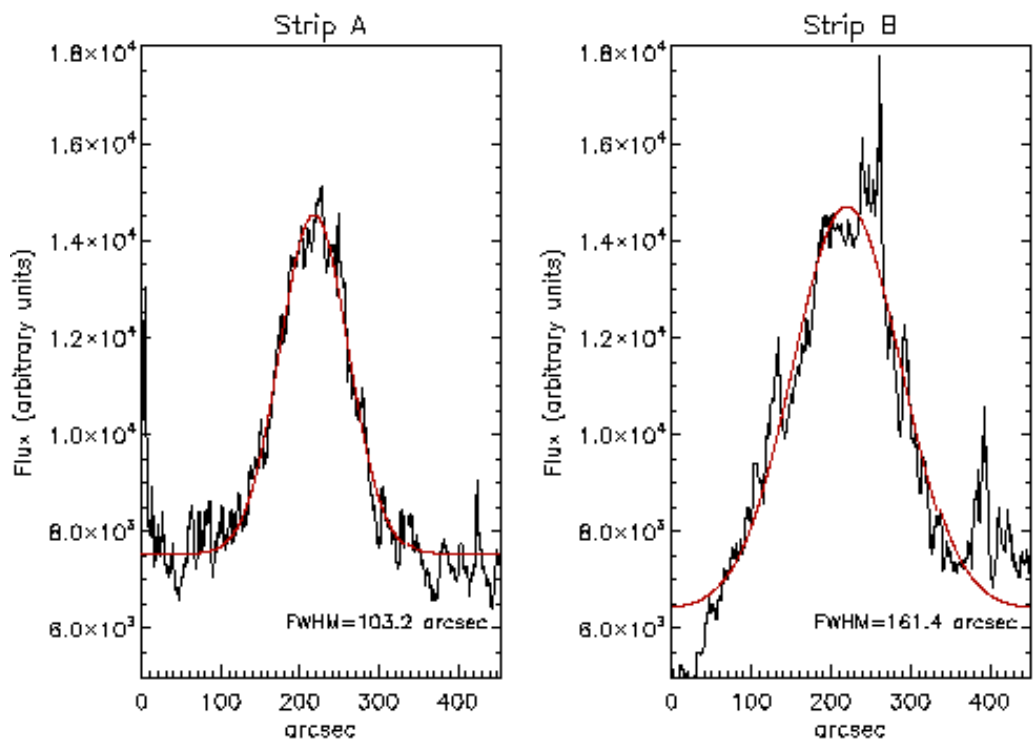}   
\caption{Gaussian fits to the surface brightness distributions of the stream
  in two different strips (A and B) illustrated in Fig.2. The FWHM of these
  Gaussian fits are estimated to be $\sim$103 and $\sim$164 arcsec
  respectively. Note that 1 arcsec at the distance of NGC5907 corresponds to
  $\sim$0.068 kpc.}
\label{width}   
\end{figure}

In order to determine the total luminosity of the tidal stream we first calculate the total area covered by
the tidal tails. To do so, we estimate the stream width in two different
strips A and B located in the bright stream piece detected by Sheng et al. (1998) (see Fig.~\ref{BBROimagePlabel}). The surface brightness
distribution in these two strips is shown in Fig.~\ref{width}.  The width of
the stream in each of these strips is estimated using a Gaussian function and the Full Width Half Maximum 
(FWHM) of these Gaussian fits are used as a
measurement of the stream width. To be conservative, we take the minimum
of the FWHM measured in both regions (i.e. $\sim$103 arcsec; see
Fig.~\ref{width}) as representative of the width of the stream. At the
distance of NGC\,5907 the above value corresponds to $\sim$7 kpc. Subsequently, we calculate the length of the stream by placing elliptical annuli at the inner at outer sides of each loop. Here we consider that the stream of Fig.~\ref{BBROimagePlabel} can be described by the loop L1 formed by the SW-SE-E1-W2 arcs and the half-loop L2 formed by the W3-E2 arcs. The area of each loop is then calculated as $A=\pi q (a^2-b^2)$, where $a$ and $b$ are the semi-major axis of the elliptical annuli and are related as $b=a-FWHM$. The minor-to-major axis ratio of the ellipses has an estimated value of $q\simeq 0.5$ for L1 and L2. We find that for L1 $a=540$ arcsec$=36.6$ kpc, so that $A_{L1}=728$ kpc$^2$, and for L2 $a'=27.5$ kpc, which implies $A_{L2}=527$ kpc$^2$. 
 On doing this, the approximate area subtended by these pieces of the stream is $A=A_{L1}+A_{L2}/2\simeq 990$ kpc$^2$.

Using this estimate of the area of the SE loop and E2 loop, the reported mean surface
 brightness of 27.7 mag arcsec$^{-2}$ by
\cite{Zheng1999} (i.e. $\sim$0.217 L$_{\sun}$ pc$^{-2}$ in the $R$-band)
implies a total $R$-band luminosity for the stream of $\sim$2$\times$10$^8$
L$_{\sun}$. This value is one order of magnitude brighter than the luminosity
reported for the Sgr dwarf galaxy ($\sim$1.9$\times$10$^7$, Majewski et al. 2003), but it
is still consistent with
the possibility that this complex structure of debris resulted from the
disruption of a single dwarf galaxy during a recent interaction with NGC\,5907
(see Sec.~\ref{comparison}). We have also roughly estimated the total stellar mass of the
stream by using the color $R-I=$0.5$\pm$0.3 given in \cite{Zheng1999}.  Using
the prescription given by \cite{Bell2001} the above color implies a
($M/L$)$_R$ $\sim1.6^{+15.0}_{-1.4}$ and consequently a total stellar mass of
$\sim$3.5$\times$10$^8$ M$_{\sun}$. This value is very uncertain due to the
large error in the stream color. So, the total stellar mass should only be
taken as a rough estimate of the stellar mass in the stream within an order of
magnitude. Given the large uncertainties of our estimate it is encouraging 
that the above stellar mass estimate is similar to the one estimated
($\sim$2$\times$10$^8$ M$_\sun$) by \cite{Johnston2001} using dynamical
considerations.

Unfortunately, none of the individual color filtered data set
were of sufficient exposure length to reveal hints of the faint surrounding
structures with enough signal--to--noise to permit detection or surface
brightness measurement. Therefore, it is not possible to constrain a
stellar population gradients
along the stream from this study.

%
\section{N-body Simulations}
\label{nbody}   
The clear picture of a repeatedly wrapped tidal stream around NGC\,5907
naturally suggests fitting the system with N-body models. Typically the number
of free parameters implemented by N-body models of satellite accretion in
external galaxies is vast. In the case of NGC\,5907, although the mass within
the luminous radius is well constrained by measurements of its circular
velocity curve, no information is available on the flattening of its potential
on the scales where the tidal remnants are observed ($\sim 10$--50 kpc in
projection). Additionally, the present position and velocity vectors of the
progenitor system are unknown as well as its mass and luminous profiles.
Currently, the only available information on the stream is its surface
brightness and projected geometry with respect to the host. This introduces a
new free parameter, which is the projection angle between the line of sight
and the orbital plane of the stream progenitor. Fortunately, NGC\,5907 presents
an edge-on disk
projection that eliminates a second projection angle between the line-of-sight
vector and the axi-symmetry plane of the disk. Kinetically data of bright
dynamical tracers (as planetary nebula and globular clusters)along the
stream  would help to
disentangle the projection effects and to constrain the satellite orbit.
Unfortunately, these cannot be readily obtained from individual stars with the present observational
capabilities owing to the large distance of NGC\,5907.

The huge degeneracy of the problem weakens any possible constraint obtained
through N-body modeling of the stream and, in fact, it is presently unclear
how much information can be gleaned from streams in external galaxies using
this technique. That is an open question that goes beyond the scope of this
paper and that will be studied in a future contribution (Pe\~narrubia et al.,
{\it in prep}).
Nevertheless, in this paper our N-body models are only used to answer a more
simple question: can the complex stellar structure be the result of a {\it
  single} merger event? This reduces the problem of finding one of the many
possible free parameter combinations that yields a reasonably good match to
the observations.
\subsection{Model Parameters}  
We examine several illustrative N-body simulations of dwarf spheroidal systems
disrupting in a spiral galaxy having properties similar to NGC\,5907. This
galaxy is an Sc spiral with a maximum rotational velocity of $V_{c,{\rm
    max}}=227$ km s$^{-1}$ and absolute magnitude $M_B\simeq -21$
\citep{Casertano1983} --- similar, therefore, to the Milky Way.
Here we assume a static host galaxy potential during the satellite orbit. As
\cite{Penarrubia2006} show, the geometry and kinematics of a tidal stream are
independent of the past evolution of the host galaxy potential and solely
reflect the present properties of the host. Our satellite galaxy realizations
are injected in a host galaxy that has the following components: a
\cite{Miyamoto1975} disk, a \cite{Hernquist1990} bulge and a Navarro,Frenk \&
White (1996,1997)
dark matter halo. The determination of the disk, bulge and dark matter halo
mass from the circular velocity curve of NGC 5907 is very uncertain due to
the well-known disk-halo degeneracy. Here we adopt values that are similar
to those derived by \cite{Just2006} by fitting the observed rotation curve:
the disk and bulge masses are  $M_d=8.4\times
10^{10}M_\odot$ and $M_b=2.3\times 10^{10}M_\odot$, respectively.  The radial
and vertical exponential scale-lengths of the disk are $R_d=6.4$ kpc and
$z_d=0.26$ kpc, respectively. 
The bulge core radius is $c=0.6$ kpc. The halo density profile can be written as
\begin{equation} 
\rho_{\rm NFW} =\frac{\delta_c\rho_c}{r/r_s (1 + r/r_s)^2}, 
\label{eq:rhonfw} 
\end{equation} 
where $\rho_c$=277.72 $h^{2}M_\odot{\rm kpc}^2$ is the present-day critical
density, $h=0.71$ is the Hubble constant in units of 100 ${\rm km}$ ${\rm
  s}^{-1}$ ${\rm Mpc}^{-1}$ consistent with constraints from
cosmic microwave background (CMB)  measurements and galaxy clustering (see Spergel et al 2007 and
references therein), $\delta_c$ is a dimensionless parameter and $r_s$
is the halo scale radius.  For $\delta_c=200$ we use $M_{200}=6.8\times
10^{11}M_\odot$, $r_s\simeq 8.2$ kpc and $r_{200}=180$ kpc. We must remark, however, that different combinations of the host 
galaxy parameters can yield a similar value of $V_{c,{\rm max}}$ (see Just
et al. 2006 for a detailed disccusion). Unfortunately, this increases the
degeneracy inherent to the models of the tidal stream surrounding NGC 5907.
The satellite galaxy is modeled as a \cite{King1966} profile for simplicity.
More realistic, multi-component models that distinguish between baryonic and
dark matter, such as those motivated by $\Lambda$CDM
\citep[e.g.][]{Penarrubia2008a,Penarrubia2008b}, would further increase the
number of free parameters.  Our satellite galaxies have typical parameters for
nearby dwarf galaxies \citep{Mateo1998}: a total mass of $2\times 10^8
M_\odot$, core radius of $R_c=390$ pc and King concentration $c_k=\log_{10}(R_t/R_c)\simeq
0.84$, where $R_t$ is the King tidal radius.  These values were selected to
reproduce the projected width and extent of the stellar tidal stream detected
in NGC\,5907 (see below).

In order to find an illustrative model that provides a reasonable match to the
projected distribution of debris, we run a set of N-body simulations that
explore some of the satellite's orbital and structural parameters. Following
\cite{Reshetnikov2000}, we assume that the orbital apocenter corresponds to
the maximum separation of the stream from the center of NGC 5907 (i.e. that of
the SE loop), so that $r_a=51$ kpc. To estimate the orbital eccentricity, we
vary the peri to apocentric distance ratio  ($r_{\rm p}/r_{\rm a}=0.1,
0.2,...,1.0$). Following the estimates of Johnston, Sackett \& Bullock (2001),
we explore three different progenitor masses ($10^8, 2\times 10^8$ and
$5\times 10^8 M_\odot$). To derive the age of the stream  we integrate each
N-body realization a maximum time of 9 Gyr, recording the positions and
velocities of the N-body particles in 70 snap-shots (i.e. with a time interval of 0.14 Gyr). Each snap-shot is then visually inspected varying the projection angle ($\xi$).

\subsection{Comparison of N-body Models to the Data} 
\label{comparison}
In Fig.~\ref{nbodyfig} we show an illustrative single-accretion model that
produces a good overall match to most of the observed structures.\footnote{This figure is the final snapshot of a movie included as supplementary material in the electronic version of this paper.} This suggests
that most of the tidal loops could have formed from the accretion of a single
dwarf satellite with an initial total mass of $2\times 10^8 M_\odot$.  For this particular model, the satellite has lost
60\% of its mass due to tidal stripping after 3.6 Gyr of evolution.  The
progenitor moves on a highly inclined orbit (orbital inclination
$i=80^\circ$ with respect to the disk plane), with relatively low eccentricity (the peri and apocenters are,
respectively, $r_p=21$ kpc and $r_a=51$ kpc, which translates into an
intermediate orbital eccentricity of $e\simeq 0.42$) and a radial orbital
period of $T_r\simeq 0.9$ Gyr. Remarkably, these parameters are 
fairly similar to those derived for the Sgr stream in the Milky Way (Law et al. 2005).
In
addition, we find that the projection angle formed by the line-of-sight and the orbital
plane vector has a present value of $\xi\approx 57^\circ$ (an edge-on view of the
stream implies $\xi=90^\circ$), which explains the considerable elongation of the
stream in the vertical direction.

In general, our model agrees with the previous model of \cite{Reshetnikov2000}, although a few differences must be noticed. For example, \cite{Reshetnikov2000} assume a perfectly polar orbit ($i=90^\circ$), which is therefore not affected by orbital precession, and derive a slightly higher orbital eccentricity ($\approx 0.6$), probably a result of the smaller amount of observational constraints (only the SE loop was known at that time).

Our model also provides a comprehensive explanation of the relation
between the different loops and arcs discussed in \S~\ref{sec:shape}. In
particular, this solution suggests that the leading arm forms the SE-loop,
bending toward the disk along the E1 loop and giving rise to the arc W2 after
it crosses the NGC\,5907 disk.  The trailing tail of the tidal stream would be
visible as the arc W1 in the west, and its continuation bends toward the east
direction to form the W3 arc and the E2 loop after it crosses the disk.  The
longest stream piece that fades away towards the west (the SW tail; Fig. 2)
would correspond to the oldest piece of the trailing tail with particles that,
according to our model, were stripped around 3.6 Gyr ago. 
Although the shape of our illustrative stream model can reproduce most of the
prominent observed features, it also predicts the continuation of the E2 loop
that extends between the SE loop and E1 arc. This feature is not obvious in
Fig.\ref{BBROimage}, although the presence of an excess of light in the area
where this arc would be located (denoted as OV in Fig.~\ref{BBROimagePlabel}) is
noticeable.

Interestingly, the model also successfully reproduces the strong east-west
stream asymmetry (the SE loop reaches a projected distance of $\simeq 51$ kpc,
whereas the SW tail only extends out to $\simeq 27$ kpc), which results from
the eccentric orbit of the progenitor system and an inclined line-of-sight projection.

 This illustrative model also provides a reasonable explanation for the fate of the progenitor satellite.  
 It suggests that the dwarf could have survived but
that its remnant core (black points in Fig.~\ref{nbodyfig}) might presently
be hidden behind the galaxy disk, $\sim15$ kpc from the galactic center.  If
this hypothesis is correct, it leads us to conclude that the NE ring
feature of \cite{Shang1998} corresponds to a young piece of the stream.
In addition, our model suggests that the NGC\,5907 stellar stream may be
relatively old, with the fainter, outer loop material (SW tail in Fig. 2) becoming unbound at
least 3.6 Gyrs ago (blue points in Fig.~\ref{nbodyfig}). The new tidal tails presented in this paper are
considerably older than the previous stream pieces discovered by Shang et al. (Reshetnikov \& Sotnikova 2000 estimate age$\simless 1.5$ Gyr) and than any definitively established piece of the Sgr tidal
stream \citep[$\leq$ 2.5 Gyrs;][]{Majewski2003,Law2005}. The stellar stream around NGC 5907 may represent therefore one
of the most ancient wraps of a tidal stream ever reported in the halo of a
spiral galaxy.

\section{Discussion}
\label{discussion} 

In this paper we present in detail an extragalactic perspective of
a stellar tidal stream surrounding the nearby spiral galaxy NGC\,5907. This
ghostly structure is an elegant example of how a single low-mass satellite
accretion occurring in the current epoch can produce a very complex,
rosette-like structure of debris dispersed in the halo of its host galaxy that
may survive for several Gigayears, a scenario predicted by N-body simulations of tidally disrupting stellar systems around the
Milky Way \citep[e.g., ][]{Law2005, Penarrubia2005, Martin2005,
  Martinez-Delgado2007}.  

The external view of the stream wrapping NGC 5907 also offers for the first time a direct
observational evidence that the material stripped from disrupting satellite galaxies (stream stars likely accompanied by dark matter) may cross the inner regions of galactic
disks. This process will likely induce gravitational effects (e.g., creation of
spiral arms, star formation triggering, 
etc) that are not yet fully understood, a phenomena that
has been also suggested for the Sgr stream in our Galaxy
\citep{Majewski2003, Martinez-Delgado2007} but that lacks confirmation due to our inner perspective.
In addition, the presence of tidal streams wrapping galaxies with warped disks (see also Mart\'inez-Delgado et al. 2008) may suggest satellite galaxy perturbations as the origin of those features (e.g. Vel\'azquez \& White 1999, Weinberg \& Blitz 2006).

Recent cosmologically-motivated simulations have shown that in a $\Lambda$CDM paradigm the accretion of several massive ($0.2$--0.6$M_{\rm disk}$) sub-halos onto the central regions of the host galaxy should be common from $z\simeq 1$ to the present (Kazantzidis et al. 2007). Such accretion events lead to strong warping, flaring and thickening of an initially cold disk as well as to the formation of long-lived, ring-like stellar features in the outskirts of the disk that may locate several kiloparsecs off the disk plane and have surface brightnesses in the range of 25--30 mag arcsec$^{-2}$ (see Fig.~6 of Kazantzidis et al. 2007). Given the similarities in shape and mass between NGC 5907 and the Milky Way, it is interesting to note that our deep images show none of the predicted features besides a classical warped disk. A statistically relevant sample of deep photometric images of nearby, edge-on spiral galaxies might reveal whether the absence of strongly perturbed old disks poses a problem for our theoretical understanding of how disk galaxies form in a $\Lambda$CDM context.

Our survey reveals that the presence of complex structures from ancient streams in the outskirts of spiral
galaxies in the Local Volume ($\simless$ 15 Mpc) must be taken into account to
correctly interpret the results of deep, pencil-beam photometric probes
designed to constrain the formation and composition of galactic halos.  This
concern is justified in the case of NGC\,5907, where several authors
\citep[e.g.,][]{Sackett1994, Lequeux1996} have reported the presence of a
peculiar, red, luminous, flattened stellar halo that may actually correspond
to tidal debris that in projection appears located closely below the disk plane
and that may have originated the faint residual light emission reported in
these previous studies \citep[as also suggested by][]{Zheng1999}.

What can we learn from the theoretical study of tidal streams in galaxies
beyond the Local Group? Firstly it must be remarked that 
theoretical models are severely hampered
by the impossibility of measuring the kinematics of individual stars
 at distances of several Mpc with present instrumentation.
In the case of NGC 5907, the large degeneracies in the orbital and structural parameters of the stream progenitor force us to consider our N-body model as illustrative only. However, the model is certainly helpful to interpret most of the observed features of the NGC 5907 tidal stream and proves that such a complex stream-like structure can be explained by the accretion of a {\it single} satellite galaxy. 

Despite the degeneracy introduced by the observational limitations, an
advantage of the panoramic
 perspective in external galaxies is that it simplifies the detection of
 stream pieces that were lost several Gigayears ago,
 which strongly increases the
number and strength of the constraints that can be derived from these systems. For
example, (i) the variation of width and surface brightness along tidal streams
constrains the number of dark matter sub-halos in the Milky Way
\citep{Ibata2002, Johnston2002, Penarrubia2006, Siegal2007}. These authors
show that the presence of dark matter sub-halos in spiral galaxies would result
in a progressive heating of tidal streams as a result of close encounters. As
\cite{Penarrubia2006} point out, the averaged number of dark matter
substructures (and, thus, the likelihood of encounters) in a Milky Way-like
galaxy decreases monotonically since $z\sim 2$ to the present, which clearly
implies that ``old'' stream pieces are more likely to reveal perturbations
than recently stripped ones. (ii) Constraints on the flattening of the host
galaxy potential become significantly stronger as the number of stream wraps
increase \citep[e.g.,][]{Johnston2001, Law2005, Penarrubia2005}. (iii) Finally,
detecting old stream pieces will allow us to study possible metallicity
gradients within the progenitor galaxy, as it is for example in the case of the
Sgr tidal stream \citep[e.g.,][]{Bellazzini2006,Chou2007}.

The detection of an old, complex tidal stream in NGC\,5907 (14 Mpc
away) with modest instruments suggests the viability of carrying out
extragalactic surveys of substructures in the surroundings of spiral
galaxies in the Local Volume. It is a relevant question to wonder how many stellar streams should we
expect in the Local Volume. Cosmological N-body simulations predict that the
incidence of large satellite galaxy accretion occurring in the current epoch
is commonly {\it one per spiral galaxy for its inner galactocentric regions},
$\simless$ 50 kpc \citep[e.g.,][]{Bullock2005}. The presence of such a complex
stream at 14 Mpc from us, easily detectable with present
observational capabilities, clearly suggests that deep photometric images may
reveal a large number of substructures in the stellar halos of spiral
galaxies. The different stellar streams detected in the Milky Way
\citep{Yanny2003, Majewski2003, Juric2005, Belokurov2006a, Belokurov2006b} and
in M31 \citep{Ibata2007} by large photometric surveys supports this
expectation.

Table~\ref{streamtable} gives the confirmed stellar stream features
discovered up to date. Interestingly, the first outstanding characteristic of 
the tidal streams discovered in external galaxies to the present day is there 
are {\it no} detected tidal streams in the Local Group with surface brightnesses 
as high as seen in NGC\,5907.

Presently, observational limitations impede the resolution of stars in external galaxies, which poses a
detection threshold of around $\mu_R\sim$ 29 \magsqarcsec in unresolved light.
In contrast, star counts can reach a detection level as low as $\sim$33 \magsqarcsec 
\citep[see e.g.,][]{Belokurov2006a}. This explains why we cannot find the counter-parts 
of the stellar streams detected in the Local Group in external galaxies, but it is indeed 
puzzling that no tidal stream in the Milky Way or M31 is remotely as bright as the one detected here.

\begin{figure}[!]   
\vspace{15.2cm}   
\includegraphics{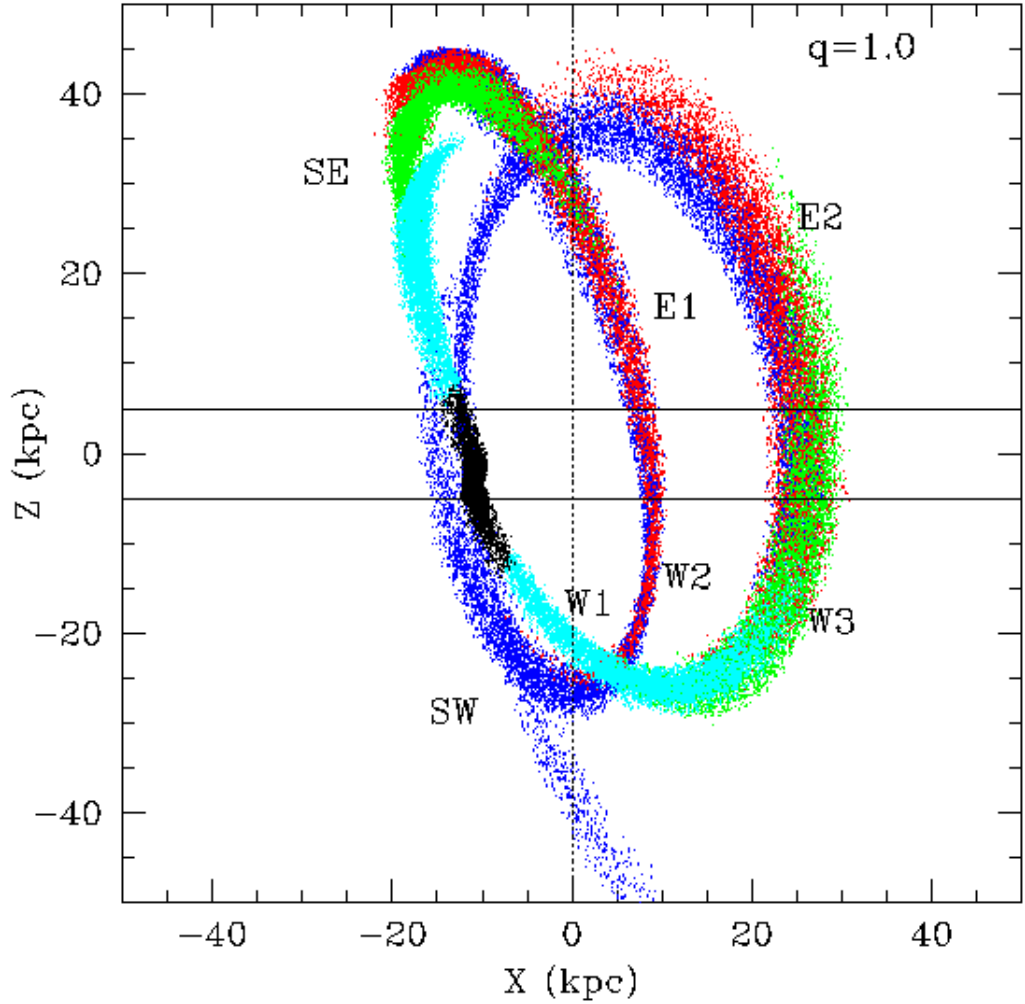}
\caption{Illustrative N-body model of the stellar stream detected in NGC
  5907. The satellite is realized as a King model with an initial mass, King
  core and tidal radii of $M=2\times10^8 M_\odot$, $r_c=0.39$ kpc and $r_t=2.7$ kpc, respectively (see Sec. 4.2 for details). Different colours denote particles that became unbound after
  different peri-center passages, whereas black particles are those that
  remain bound to the main system. Our color code is so that cyan, green red
  and blue colours are particles lost during the first, second, third and four
  orbital periods, respectively. For this particular model the orbital period
  is $T_r=0.9$ Gyr. }
\label{nbodyfig}   
\end{figure}

In addition, Table~\ref{streamtable} also reveals that, with the single
exception of the Sgr dwarf, the progenitor systems of most of the stellar
streams known to the present day remain {\it undetected}.  The most logical
explanation is that the progenitor galaxies have been nearly destroyed by
now. This scenario is supported by the N-body simulations of
\cite{Penarrubia2008b}, who show that dwarf galaxies undergoing tidal mass
stripping have to lose more than 90--99\% of their initial dark matter halo
mass before a stellar tidal stream may start forming. Once the host's tidal
field begins to strip stellar material, the central surface brightness
($\Sigma_0$) and the total luminosity ($L$) of the dwarf galaxy suffers a
sharp, monotonic decline so that $\Sigma_0/\Sigma_0(t=0)\propto
\big[L/L(t=0)\big]^{0.6}$ for $L/L[t=0]\ll 1$.
Therefore, as a tidal stream forms the likelihood to detect its progenitor
decreases until, eventually, the present surface brightness falls beyond the
detection limits of our instruments.  A second possibility that we explore in
our illustrative model (see Fig.~\ref{nbodyfig}) is that the progenitor
systems might be hidden behind the host disk or bulge. Although this scenario
may be applied to some of the systems listed in Table~\ref{streamtable}, it is
fairly unlikely that it explains the absence of progenitor galaxies for all of
them.

The study of tidal streams in external galaxies is a relatively young field
and provides direct way of addressing some opened questions on galactic 
formation and evolution. For example, studying stellar
population gradients along tidal streams via deep {\it Hubble Space Telescope}
photometry data
\citep[see for example][]{Aloisi2005} will render important
constraints on the effect of tides on the stellar formation history of
 dwarf galaxies. The panoramic view of tidal streams in external galaxies also 
 offers an excellent opportunity to  demonstrate tidal stripping of globular
clusters formed in satellite galaxies, which may correspond to an
important fraction of the globular cluster population of the host, as
earlier proposed by \cite{Searle1978}. Searching for associated globular
clusters and planetary nebula may also offer a means by which kinematics
in distant streams could be derived (i.e. from intrinsically bright absorption line sources
or sources with emission lines), so that kinematical constraints might be placed
on theoretical models that aim to reproduce tidal streams in external
galaxies.  These models, in turn, may provide reasonable estimates of the
flattening of the dark matter halos that surround Local Volume
galaxies. 
Ultimately, the ideal scenario would require resolving
stellar populations at a distance of 10--20 Mpc, which will be feasible in the
next one or two decades with large ground-based telescopes like the
{\it Thirty-Meter-Telescope}, or spatial missions like the {\it James Webb Space
Telescope}.

\vskip1cm
We thank  H.-W. Rix, E. Bell, A. McConnachie, E. J. Alfaro, R. Ibata and
L. Mayer for very fruitful discussions. We are also grateful to the anonymous 
referee for constructive suggestions that helped to improve this manuscript.
 D. M.-D. acknowledges the hospitality of the Max
Planck Institute for Astronomy (Heidelberg, Germany), the Institute for
Advance Studies (Princeton, USA) and the University of Victoria (Canada), where part of this work was done. D. M-D
acknowledge funding from the Spanish Ministry of Education and Science (Ramon
y Cajal program contract and research project AYA 2007-65090).
 JP
thanks Julio F. Navarro for financial support.  SRM acknowledge funding by NSF grant 
AST-0307851 and NASA/JPL contract 1228235.
\begin{deluxetable}{lclc} 
\tablecaption{JOURNAL OF OBSERVATIONS}  
\tablewidth{12cm} 
\tablehead{\colhead{Date} & & \colhead{Filter} & \colhead{Total Exposure Time (s)}} 
\startdata 
2006 June 6  & &Blue &  6480\\ 
2006 June 6   & & Clear Luminance & 9000\\ 
2006 June 6   & & Green & 1080\\ 
2006 June 6   &  &Red &  5400\\ 
2006 June 7   &  &Clear Luminance &  3600\\ 
2006 June 7   &   &Green & 2160\\ 
2006 June 9   & &Clear Luminance &  5400\\ 
2006 August 10  & & Blue &  2160 \\  
2006 August 10  & & Clear Luminance &  2700\\ 
2006 August 10  & & Green &  1080\\ 
2006 August 10  &  & Red &  1800 \\ 
\hline 
\enddata 
\label{obslog} 
\end{deluxetable} 

\begin{table}
\caption{Stellar tidal streams detected in nearby spiral galaxies ($D<15$ Mpc)}
\begin{tabular}{l  c  c c c  } \hline \hline
Name & Distance (Mpc)  & $\mu$ (band) & Progenitor & Refs. \\ \hline
{\bf Local Group} &&&& \\ \hline \hline
Sagittarius (MW) & --  & 29.6 (V)  & Sagittarius dwarf &(1)  \\
Monoceros (MW)   & --  & 34.9 (V)  & Canis Major dwarf (?) &(2),(3) \\
Orphan (MW)    & --  & 32.4 (R)  & unknown&(4) \\
M31 giant stream (M31) & 0.78 & 30.0 (V) & unknown &(5),(6) \\
{\bf External galaxies} &&& \\ \hline \hline
NGC 5236 (M83) & 5.2 & 26.5 (R) & unknown &(7),(8) \\
NGC 5907       & 11.2& 26.8 (R) & unknown &(9),(10)  \\
NGC 4013       & 12.0& 27.0 (R) & unknown &(11)  \\
NGC 4594 (M104)& 12.4& --       & unknown &(8)\\
NGC 3310       & 14.4& 25.6 (V) & unknown &(14)  \\
\hline \hline
\end{tabular}\label{streamtable}
\tablerefs{(1) Mart\'inez-Delgado et al. (2004); (2) Belokurov et al. (2006); (3) Martin et al. (2004); (4) Belokurov et al. (2007); (5) Ibata et al.
(2001);(6) McConnachie et al. (2003); (7) de Jong et al. (2207);(8) Malin \& Hadley (1997); (9)
Shang et al. (1998); (10) this manuscript; (11) Mart\'inez-Delgado et al. (2008); (12) Whener
\& Gallagher (2005)}

\end{table}


\begin{thebibliography}{}
%
%
\bibitem[Aloisi et al.(2005)]{Aloisi2005} Aloisi, A., van der Marel, R. P.,
  Mack, J., Leitherer, C., Sirianni, M., Tosi, M. 2005, \apj, 631, L45
%
\bibitem[Barnaby \& Thronson(1992)]{Barnaby1992} Barnaby, D., \& Thronson, H.~A., Jr.\ 1992, \aj, 103, 41
%
\bibitem[Bell \& de Jong(2001)]{Bell2001} Bell, E. \& de Jong, R. S. 2001, \apj, 550, 212
%
\bibitem[Bellazzini et al.(2006)]{Bellazzini2006} Bellazzini, M., Newberg, H.~J., Correnti, M., Ferraro, F.~R., \& Monaco, L.\ 2006, \aap, 457, L21 
%
\bibitem[Belokurov et al.(2006a)]{Belokurov2006a} Belokurov, V., et  
al.\ 2006a, \apjl, 642, L137  
%
\bibitem[Belokurov et al.(2006b)]{Belokurov2006b} Belokurov, V., Evans,  
N.~W., Irwin, M.~J., Hewett, P.~C., \& Wilkinson, M.~I.\ 2006b, \apjl, 637,  
L29  
%
\bibitem[Biermann \& Munyaneza(2007)]{Biermann2007} Biermann, P.~L.,  
\& Munyaneza, F.\ 2007, ArXiv Astrophysics e-prints, arXiv:astro-ph/0702173 
%
\bibitem[Bullock \& Johnston(2005)]{Bullock2005} Bullock, J.S., Johnston, K. V. 2005, \apj, 635, 931   
%
\bibitem[Casertano(1983)]{Casertano1983} Casertano, S.\ 1983, \mnras, 203, 735
%
\bibitem[Chou et al.(2007)]{Chou2007} Chou, M.-Y., et al.\ 2007, 
\apj, 670, 346
%
\bibitem[Dave et al.(2001)]{Dave2001} Dav\'e, R., Spergel, D. N., Steinhardt,
  P. J., Wandelt, B. D.\ 2001, 
\apj, 547, 574

\bibitem[Davies(1990)]{Davies1990} Davies, E. 1990, in {\it Machine Vision: Theory, Algorithms and Practicalities}, Academic Press, pp. 42
%

\bibitem[Dubinsky et al.(1991)]{Dubinsky2001} Dubinsky, J. \& Calberg, R. G.\ 1991, \apj, 378, 496

\bibitem[Dumke et al.(1997)]{Dumke1997} Dumke, M., Braine, J., Krause, M., Zylka, R., Wielebinski, R., \& Guelin, M.\ 1997, \aap, 325, 124   
%
\bibitem[Erwin, Pohlen, \& Beckman(2008)Erwin et al.]{Erwin2008} Erwin, P., Pohlen, M., \& Beckman, J.E.\ 2008, AJ, 135, 20
%
\bibitem[Florido et al.(1992)]{Florido1992} Florido, E., Battaner, E., Gros, A., Prieto, M., \& Mediavilla, E.\ 1992, \apss, 190, 293  
%
\bibitem[Forbes et al.(2003)]{Forbes2003} Forbes, D. A., Beasley, M. A., Bekki, K., Brodie, J. P., Strader, J. 2003, Science, 301, 1217   

\bibitem[Haralick \& Shapiro(1992)]{Haralick1992} Haralick, R., Shapiro, L. 1992, in {\it Computer and Robot Vision}, Addison-Wesley Publishing Company, volume 1, chapter 7.  
%
\bibitem[Hayashi, Navarro, \& Springel(2007)Hayashi et al.]{Hayashi2007}
  Hayashi, E., Navarro, J.~F., \& Springel, V.\ 2007, \mnras, 377, 50
%
\bibitem[Helmi(2004)]{Helmi04} Helmi, A.\ 2004, \apjl, 610, L97 
%
\bibitem[Hernquist(1990)]{Hernquist1990} Hernquist L., 1990, ApJ, 356, 359  
%
\bibitem[Ibata \& Lewis(1998)]{Ibata1998} Ibata, R.~A., \& Lewis, G.~F.\ 1998,
  \apj, 500, 575
%
\bibitem[Ibata et al.(2001a)]{Ibata2001a} Ibata, R., Irwin, M., Lewis, G., Ferguson, A.~M.~N., \& Tanvir, N.\ 2001a, Nature, 412, 49 
%
\bibitem[Ibata et al.(2001b)]{Ibata2001b} Ibata, R., Lewis, G.~F., Irwin, M., Totten, E., \& Quinn, T.\ 2001b, \apj, 551, 294   
%
\bibitem[Ibata et al.(2002)]{Ibata2002} Ibata, R.~A., Lewis, G.~F., Irwin,
  M.~J., \& Quinn, T.\ 2002, \mnras, 332, 915
%
\bibitem[Ibata et al.(2007)]{Ibata2007} Ibata, R., Martin, N.~F., Irwin, M.,
  Chapman, S., Ferguson, A.~M.~N., Lewis, G.~F., \& McConnachie, A.~W.\ 2007,
  \apj, 671, 1591
%
\bibitem[Irwin \& Madden(2006)]{Irwin2006} Irwin, J.~A., \& Madden, S.~C.\ 2006, \aap, 445, 123   
%
\bibitem[James \& Casali(1998)]{James1998} James, P.~A., \& Casali, M.~M.\ 1998, \mnras, 301, 280  
%
\bibitem[Johnston, Spergel, \& Hernquist(1995)Johnston et al.]{Johnston1995} Johnston, K.~V., Spergel, D.~N., \& Hernquist, L.\ 1995, \apj, 451, 598   
%
\bibitem[Johnston, Sackett, \& Bullock(2001)Johnston et al.]{Johnston2001} Johnston, K.~V., Sackett, P.~D., \& Bullock, J.~S.\ 2001, \apj, 557, 137   
%
\bibitem[Johnston, Choi, \& Guhathakurta(2002)Johnston et al.]{Johnston2002}
  Johnston, K.~V., Choi, P.~I., \& Guhathakurta, P.\ 2002, \aj, 124, 127
%
\bibitem[Johnston et al.(2005)]{Johnston2005} Johnston, K. V., Law, D. R., Majewski, S. R. 2005, \apj, 619, 800
%
\bibitem[Juric et al.(2005)]{Juric2005} Juric, M., et al.\ 2008, \apj, 673, 864 
\bibitem[Just et al.(2006)]{Just2006} Just, A., M{\"o}llenhoff, C., \& Borch, A.\ 2006, \aap, 459, 703  



%
\bibitem[Kazantzidis et al.(2004)]{Kazantzidis2004} Kazantzidis, S.,  
Kravtsov, A.~V., Zentner, A.~R., Allgood, B., Nagai, D., \& Moore, B.\  
2004, \apjl, 611, L73  
%
\bibitem[Kazantzidis et al.(2007)]{2007arXiv0708.1949K} Kazantzidis, S., 
Bullock, J.~S., Zentner, A.~R., Kravtsov, A.~V., 
\& Moustakas, L.~A.\ 2007, ArXiv e-prints, 708, arXiv:0708.1949 
%
\bibitem[King(1966)]{King1966} King, I.~R.\ 1966, \aj, 71, 64   
%
\bibitem[Law, Johnston, \& Majewski(2005)Law et al.]{Law2005} Law, D. R.,
  Johnston, K. V., \& Majewski, S. R. 2005, \apj, 619, 807
%
\bibitem[Lequeux(1996)]{Lequeux1996} Lequeux, J., Fort, B., Dantel-Fort, M., Cuillandre, J.-C., \& Mellier, Y.\ 1996, \aap, 312, L1  
%
\bibitem[Lequeux(1998)]{Lequeux1998} Lequeux, J., Combes, F., Dantel-Fort, M., Cuillandre, J.-C., Fort, B., \& Mellier, Y.\ 1998, \aap, 334, L9    
%
\bibitem[Majewski et al.(2003)]{Majewski2003} Majewski, S. R., Skrutskie, M., Weinberg, M., Ostheimer, J. 2003, \apj, 599, 1082   
%
\bibitem[Malin \& Hadley(1997)]{Malin1997} Malin, D. \& Hadley, B. 1997, PASA, 14,52   
%
%
\bibitem[Martin et al.(2005)]{Martin2005} Martin, N.~F., Ibata, R.~A., Conn,
  B.~C., Lewis, G.~F., Bellazzini, M., \& Irwin, M.~J.\ 2005, \mnras, 362, 906
%
\bibitem[Mart\'\i nez-Delgado et al.(2004)]{Martinez-Delgado2004} Mart\'\i
  nez-Delgado, D.,  G\'omez-Flechoso, M. A., Aparicio, A., \& Carrera, R. 2004, \apj, 601, 242   
%
\bibitem[Mart\'\i nez-Delgado et al.(2007)]{Martinez-Delgado2007} Mart\'\i
  nez-Delgado, D., Pe\~narrubia, J., Juric, M., Alfaro, E. J., Ivezic,
  Z. 2007, \apj, 660,1264
%
\bibitem[Mart\'\i nez-Delgado et al.(2008)]{Martinez-Delgado2008} Mart\'\i
  nez-Delgado, D., Pohlen, M., Gabany, R. J., Majewski, S.~R., Pe\~narrubia, J., \& Palma, C. 2008, \apj, submitted 
%
\bibitem[Mateo(1998)]{Mateo1998} Mateo, M.~L.\ 1998, \araa, 36, 435 
%
\bibitem[McConnachie et al.(2003)]{McConnachie2003} McConnachie, A. W., Irwin, M. J., Ibata, R. A., Ferguson, A. M. N., Lewis, G. F., Tanvir, N.,\mnras, 343, 1335 
%
\bibitem[Miyamoto \& Nagai(1975)]{Miyamoto1975}Miyamoto M., Nagai R., 1975, Publ. Astron. Soc. Japan, 27, 533  
%
\bibitem[Morrison, Boroson, \& Harding(1994)Morrison et al.]{Morrison1994} Morrison, H.~L., Boroson, T.~A., \& Harding, P.\ 1994, \aj, 108, 1191   
%
\bibitem[Navarro, Frenk, \& White(1996)Navarro et al.]{NFW1996} Navarro J.,
  Frenk C.S., \& White S.D.M., 1996, ApJ, 462, 563 

\bibitem[Navarro, Frenk, \& White(1997)Navarro et al.]{NFW1997} Navarro J.,
  Frenk C.S., \& White S.D.M., 1997, ApJ, 490, 493  
%
 
%
\bibitem[Pe{\~n}arrubia et al.(2005)]{Penarrubia2005} Pe\~narrubia et al. 2005, \apj, 626, 128   
%
\bibitem[Pe{\~n}arrubia et al.(2006)]{Penarrubia2006} Pe{\~n}arrubia, J.,
  Benson, A.~J., Mart{\'{\i}}nez-Delgado, D., \& Rix, H.~W.\ 2006, \apj, 645,
  240
%
\bibitem[Pe{\~n}arrubia et al.(2008a)]{Penarrubia2008a} Pe{\~n}arrubia, 
J., \& McConnachie, A.~W., Navarro, J.~F.,\ 2008a, \apj, 672, 904
%
\bibitem[Pe{\~n}arrubia et al.(2008b)]{Penarrubia2008b} Pe{\~n}arrubia, 
J., Navarro, J.~F., \& McConnachie, A.~W.\ 2008b, \apj, 673, 226 
%
\bibitem[Pohlen et al.(2004)]{Pohlen2004} Pohlen, M., Mart\'\i nez-Delgado, D., Majewski, S. R, Palma, C., Prada, F., Balcells, M. 2004, in Satellite and Tidal Streams, PASP, vol 327, pg. 288   
%
\bibitem[Pohlen \& Trujillo(2006)]{Pohlen2006} Pohlen, M., \& Trujillo, I.\ 2006, A\&A, 454, 759  
%
\bibitem[Rand(1996)]{Rand1996} Rand, R.~J.\ 1996, \apj, 462, 712   
%
\bibitem[Reshetnikov \& Sotnikova(2000)]{Reshetnikov2000}
  Reshetnikov, V. P., \& Sotnikova, N. Y. 2000, Astronomy Letters, Vol. 26, 277 %
\bibitem[Rudy et al.(1997)]{Rudy1997} Rudy, R.~J., Woodward, C.~E., Hodge, T., Fairfield, S.~W., \& Harker, D.~E.\ 1997, \nat, 387, 159   
%
\bibitem[Sackett et al.(1994)]{Sackett1994} Sackett, P.~D., Morrison, H.~L., Harding, P., \& Boroson, T.~A.\ 1994, \nat, 370, 441   
%
\bibitem[Sancisi(1976)]{Sancisi1976} Sancisi, R.\ 1976, \aap, 53, 159 
%
\bibitem[Sasaki(1987)]{Sasaki1987} Sasaki, T.\ 1987, \pasj, 39, 849   
%
\bibitem[Searle \& Zinn(1978)]{Searle1978} Searle, L., Zinn, R. 1978, \apj, 225, 357   
%
\bibitem[Shang et al.(1998)]{Shang1998} Shang, Z., et al.\ 1998, \apjl, 504, L23%
\bibitem[Siegal-Gaskins \& Valluri(2007)]{Siegal2007} Siegal-Gaskins,
  J.~M., \& Valluri, M.\ 2007, ArXiv e-prints, 710, arXiv:0710.0385 
%
\bibitem[Skrutskie, Shure, \& Beckwith(1985)Skrutskie et al.]{Skrutskie1985} Skrutskie, M.~F., Shure, M.~A., \& Beckwith, S.\ 1985, \apj, 299, 303  
%
\bibitem[Spergel et al.(2007)]{2007ApJS..170..377S} Spergel, D.~N., et al.\ 
2007, \apjs, 170, 377 
%
%
\bibitem[van der Kruit \& Searle (1981)]{vdK1981} van der Kruit, P.~C., \& Searle, L.\ 1981, \aap, 95, 105   
%
\bibitem[Velazquez \& White(1999)]{Velazquez1999} Velazquez, H. \& White, S. D. M. 1999, \mnras, 304, 254
%
\bibitem[Weinberg \& Blitz(2006)]{Weinberg2006} Weinberg, M.~D., \&  
Blitz, L.\ 2006, \apjl, 641, L33  
%
\bibitem[Xilouris et al.(1999)]{Xilouris1999} Xilouris, E.~M., Byun, Y.~I., Kylafis, N.~D., Paleologou, E.~V.,  \& Papamastorakis, J.\ 1999, \aap, 344, 868  
%
\bibitem[Yanny et al.(2003)]{Yanny2003} Yanny, B., et al.\ 2003, \apj, 588,
  824 

\bibitem[Yoshida et al.(2000)]{Yoshida2000} Yoshida, H., Springel, V., White,
  S. D. M., \& Tormen, G. 2000, \apj, 535, L103

%
\bibitem[Zepf et al.(2000)]{Zepf2000} Zepf, S. E., Liu, M.C., Marleau, F. R., Sackett, P. D. \& Graham, J. R. 2000, \aj, 119, 1701   
%
\bibitem[Zheng et al.(1999)]{Zheng1999} Zheng, Z., et al.\ 1999, \aj, 117, 2757
%
\end{thebibliography}
\end{document}